\documentclass[prl,twocolumn,aps]{revtex4}
\usepackage{amsmath}
\usepackage{amssymb}
\usepackage{graphicx}
\usepackage{epstopdf}

\begin{document}

\title{Quantum Statistics of Two Identical Particles and Modified Hong-Ou-Mandel Interferometer}
%From Quantum Indistinguishability to Classical Distinguishability: Statistics and Measurement of Two-Particle Distributions}

\author{Won-Young Hwang}
\email{wyhwang@jnu.ac.kr}
\affiliation{Department of Physics Education, Chonnam National University, Gwangju 61186, Republic of Korea}

\author{Kicheon Kang}
\email{kckang@jnu.ac.kr}
\affiliation{Department of Physics, Chonnam National University, Gwangju 61186, Republic of Korea}

\date{\today}

\begin{abstract}
%{\sf SHould be revised}\\
We propose an experimental scheme to probe the quantum statistics of two identical particles. The transition between the quantum and classical statistics of two identical particles is described by the particles having identical multiple internal energy levels. We show that effective distinguishability emerges as the thermal energy increases with respect to the energy level spacing, and the mesoscopic regime bridges quantum indistinguishability and classical distinguishability. A realistic experimental approach is proposed using a two-particle interferometer, where the particles reach statistical equilibrium before the two-particle distribution is measured. The unitarity of the scattering/separation process ensures the preservation of the equilibrium distribution and allows a direct measurement of the two-particle statistical distribution. Our results show the transition between quantum and classical behavior of the two-particle distribution, which can be directly probed by a realistic experiment.
\end{abstract}

\maketitle

%\section{Introduction}
{\em Introduction-}.
%{\sf cite Feynman where? Revise the introduction}\\
The indistinguishability of identical particles plays the key role in quantum statistics, which differs from the usual classical statistics~(see, for exmaple, Ref.~\onlinecite{feynman3-65}). An elementary but remarkable example is two-particle quantum statistics. 
Consider two identical bosons, each with a single possible state. The indistinguishability of the particles leads to three possible states of the two-particle system, in sharp contrast to the four possible states of two distinguishable particles. In the usual approach to particle statistics, there are two different ways of looking at the identical particles, namely quantum indistinguishability and ``classical" distinguishability. However, this dichotomy is incomplete because there is no such thing as a classical (distinguishable) particle in the framework of quantum theory. In other words, the emergence of classical statistics of distinguishable particles should be described in terms of quantum theory.

In this Letter, we show that the transition between quantum and classical statistics of two particles can be described within the framework of quantum theory. The essence is that each particle can have several internal levels, and the two identical particles acquire effective distinguishability as the effective number of levels increases. 
More importantly, we present an experimental scheme to measure the mesoscopic two-particle statistics, i.e. the transition from quantum to classical statistics, using two-particle interferometers for the equilibrated two particles.  The two-particle interferometer considered here is similar to the Hong-Ou-Mandel (HOM) interferometer~\cite{hong87}, but the point is that the particles should be equilibrated before the statistical distribution is measured.  It is shown that the two-particle distribution can be probed directly with this type of interferometer. The unitarity of the two-particle scattering or separation process plays an essential role in this measurement. The transition from the pure quantum to the classical limit is described, which is determined by the ratio between the thermal energy and the level spacing of each particle. 
% \\ {\sf MORE To be discussed in the INTRODUCTION?}

%\section{Quantum statistics of two identical particles}
{\em Quantum statistics of two identical particles-}.
We begin with a pedagogical argument about how classical behavior emerges from the quantum statistics of two identical particles. 
Each particle can be in two possible states (sites), $\lvert p\rangle$ and 
$\lvert q\rangle$. The composite system evolves over time in the environment 
and reaches a statistical equilibrium in which the system has an equal 
probability of occupying each of the available states. 
For distinguishable particles, there are four possible states. 
In contrast, quantum indistinguishability leads to different statistics. 
For bosons, there are three possible states; $\lvert pp\rangle$, 
$\lvert qq\rangle$, and $(\lvert pq\rangle + \lvert qp\rangle)/\sqrt{2}$. 
For fermions, there is only one possible state, 
$(\lvert pq\rangle - \lvert qp\rangle)/\sqrt{2}$, due to the Pauli principle. 
These statistical properties can be found in the probability of the two particles being in two different locations, $P(1,1)$.
For classical particles, bosons, and fermions, this probability is $1/2$, $1/3$, and $1$, respectively. 

All matter obeys the laws of quantum mechanics, and within the framework of quantum theory there is no intrinsic distinction for two identical particles if their wave functions overlap. The emergence of classical distinguishability should be understood within a single framework. An essential property of real particles is that each particle can have several different internal states, represented by $\varepsilon_{n}$ (Fig. 1). For simplicity, we assume an equal spacing of the energy levels, denoted by $\Delta$, for each particle. 
At low temperatures ($T$), only the ground state of each particle contributes to the statistics ($kT \ll \Delta$, where $k$ is the Boltzmann constant). Therefore, in this limit, the probability $P(1,1)$ for two identical bosons (fermions) reduces to $1/3$ ($1$).
%\begin{figure}[l]
\begin{figure}
\centering
\includegraphics[width=7cm]{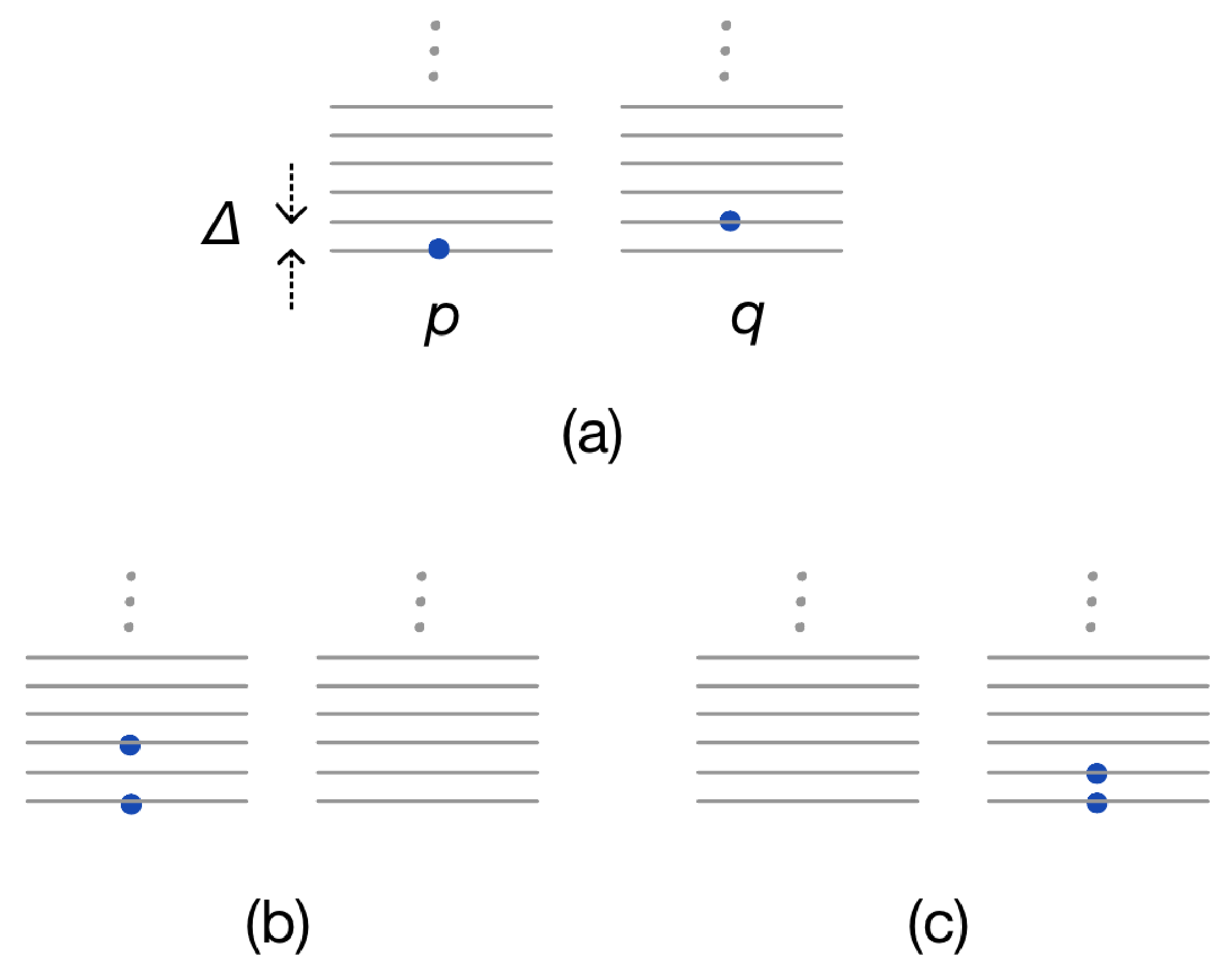} 
\caption{Possible configurations of two identical particles, each with internal energy levels (spacing $\Delta$).
 (a) Two identical particles occupying distinct sites ($p$ and $q$). (b) (c) Both particles localized at the same site.
At low temperatures ($kT\ll\Delta$), quantum statistics dominate; at high temperatures  ($kT\gg \Delta$), thermal excitations lead to effective distinguishability.
 }
\end{figure}

At higher temperatures, however, excited states come into play. It is more likely that the two particles are in different internal states and they become effectively distinguishable. As a result, the probability $P(1,1)$ is expected to be $1/2$ in the $kT \gg \Delta$ limit, for both bosons and fermions. An interesting point is that there is a mesoscopic region connecting the two limits at the intermediate temperature ($kT \sim \Delta$), where the two particles are partially distinguishable. Therefore, we expect $P(1,1)$ to show a transition between pure quantum and classical behavior.

The two-particle distribution can be calculated exactly as follows. The partition function ($Z_{2}$) of the two identical particles is given by
\begin{equation}
Z_{2} = \text{Tr}\, e^{-\beta H},
\end{equation}
where $H$ is the Hamiltonian of the two-particle system. It can be classified as
\begin{subequations}
\begin{equation}
Z_{2} = Z_{pq} + Z_{p} + Z_{q},
\end{equation}
where $Z_{pq}$ is the contribution of two particles at each position $p$ and $q$, and $Z_{p}$ ($Z_{q}$) represents the two particles at the same position $p$ ($q$). For both bosons and fermions,
\begin{equation}
Z_{pq} = \left(\sum_{n} e^{-\beta \varepsilon_{n}}\right)^{2}.
\end{equation}
$Z_{p}$ ($Z_{q}$) contains different statistics for bosons and fermions. We find
\begin{equation}
 Z_{p} = Z_{q} = \frac{1}{2} \sum_{m \neq n} 
   e^{-\beta (\varepsilon_{m} + \varepsilon_{n})} 
   + \sum_{n} e^{-2 \beta \varepsilon_{n}}
\label{eq:ZA}
\end{equation}
\end{subequations}
for two identical bosons. For fermions, the 2nd term of the RHS of 
Eq.~\eqref{eq:ZA} is not present because of the Pauli principle.

For the energy levels with equal separation $\Delta$, we obtain the analytical expression
\begin{subequations}
\label{eq:P11}
\begin{eqnarray}
 P(1,1) &=& \frac{e^{\beta \Delta} + 1}{3 e^{\beta \Delta} + 1} 
   \quad \text{(bosons)}, \\
 P(1,1) &=& \frac{e^{\beta \Delta} + 1}{e^{\beta \Delta} + 3} 
   \quad \text{(fermions)}.
\end{eqnarray}
\end{subequations}
These analytical results capture the full temperature dependence of the quantum-to-classical transition.
Fig. 2 shows the behavior of $P(1,1)$ as a function of $kT/\Delta$. At low temperatures ($kT/\Delta \ll 1$), it reduces to $1/3$ and $1$, the ideal values for identical bosons and fermions, respectively. As the temperature increases, excited states come into play, and at $kT/\Delta \gg 1$, $P(1,1)$ reaches the classical value of $1/2$ for both bosons and fermions.
\begin{figure}
\centering
\includegraphics[width=7cm]{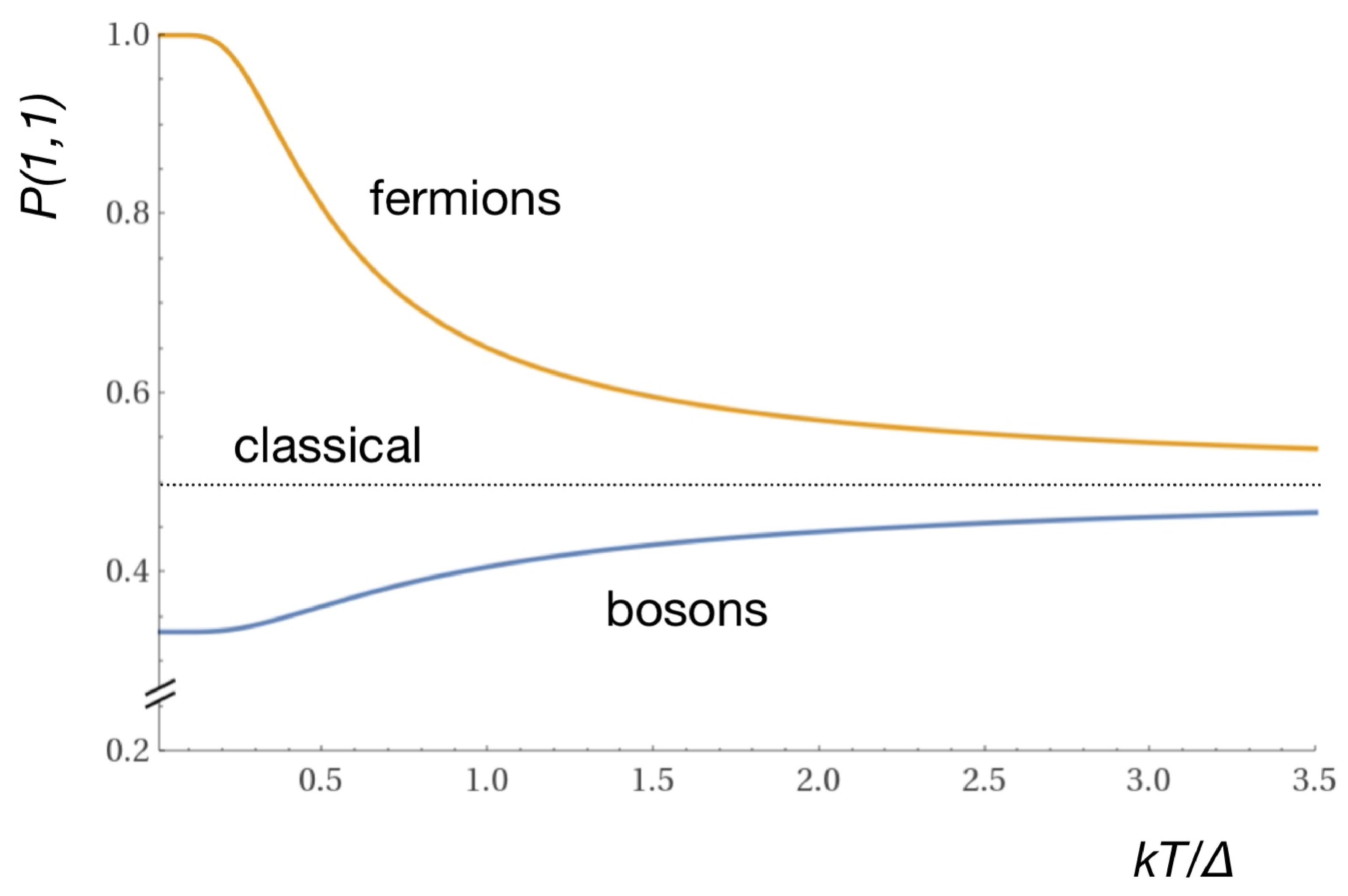} 
\caption{Transition of the two-particle distribution $P(1,1)$ as a function of $kT/\Delta$. Solid curves
show results for both bosons (lower) and fermions (upper), demonstrating the smooth transition between 
quantum statistics ($P(1,1) = 1/3$ for bosons and 1 for fermions at  $kT \ll \Delta$) and classical statistics
 ($P(1,1) = 1/2$ for both at $kT \gg \Delta$).
 }
\end{figure}

It should be noted that the two-particle distribution of Eq.~\eqref{eq:P11} is
different from the standard Bose-Einstein (BE) and Fermi-Dirac (FD) 
distributions, although they are closely related. 
In general, the statistical distributions of identical bosons and fermions 
are determined by bunching and antibunching properties in exchange scattering, 
respectively~\cite{feynman3-65}.
The BE and FD distributions represent the single particle probability 
at equilibrium in the presence of many identical particles. On the other hand, 
the two-particle distribution discussed here is applied when there are only two 
identical particles.  

%\section{A modified Hong-Ou-Mandel interferometer}
{\em Two-particle interferometer for the measurement of equilibrium statistical distributions-}.
A realistic implementation of the measurement can be achieved by two-particle interference similar to the HOM effect. The original HOM interferometer probes the interference in the exchange scattering of two identical particles. Our proposal is to use a modified version of the HOM interferometer as shown in Fig. 3. The proposed measurement procedure is as follows. First, each particle is injected from two separate input ports, $p$ and $q$. Then, the two particles are controlled to reach statistical equilibrium. This is the step not included in the original HOM interferometer. Finally,  the distribution (coincidence) of the two particles is measured at the output ports.
\begin{figure}
\centering
\includegraphics[width=7cm]{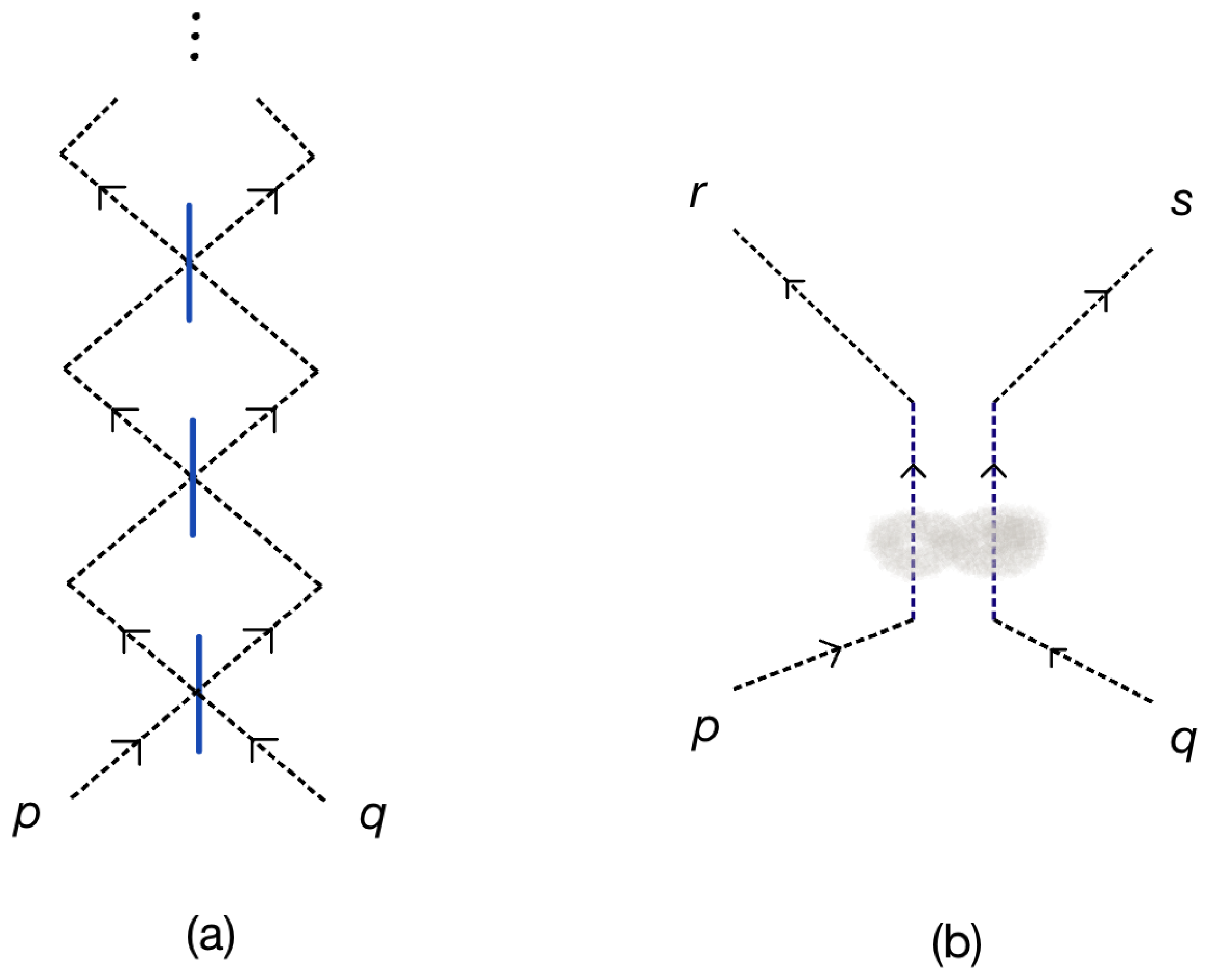} 
\caption{Proposed experimental schemes for the measurement of two-particle distribution. (a) Beam-splitter array with dephasing drives the system to equilibrium.
(b) Equilibration by controlled interaction (e.g. for cold atoms), followed by unitary separation to measure the distribution.
 }
\end{figure}

In the first example (Fig.~3(a)), we consider a series of beam splitters 
with dephasing.
The scattering at the beam splitter is described by
\begin{subequations}
\begin{equation}
\begin{pmatrix}
 c_{pn}' \\ c_{qn}'
\end{pmatrix}
 = S
\begin{pmatrix}
 c_{pn} \\ c_{qn}
\end{pmatrix},
\end{equation}
where
\begin{equation}
S = \begin{pmatrix}
r & t' \\
t & r'
\end{pmatrix} ,
\end{equation}
\end{subequations}
and $c_{\alpha n}$ 
($c_{\alpha n}'$) 
 is an annihilation operator for the particle from the input (output) 
port $\alpha$ ($= p,q$) and internal state $n$. 
The different statistics for bosons and fermions are included in the commutation relations of $c_{\alpha n}$.
Each element of the scattering matrix $S$ represents a reflection ($r,r'$) or transmission ($t,t'$) at the beam splitter.
Let us first consider the $T = 0$ case of identical bosons in the strong 
dephasing limit, i.e. the off-diagonal terms in the density matrix are discarded. Note that this can be easily generalized to the case of finite temperature. The scattering process at the $i$-th beam splitter with subsequent dephasing is described by the recurrence relation of the density matrix $\rho_i$:
\begin{subequations}
\label{eq:rhoi}
\begin{equation}
\rho_{i} = \text{Diag}[S_{T} \rho_{i-1} S_{T}^{\dagger}],
\end{equation}
where 
\begin{equation}
 S_{T} = S \otimes S
\label{eq:ST}
\end{equation}
represents the scattering of two-particles at each beam splitter.
In the strong dephasing limit, the coherence between the different paths of the particles quickly disappears, and the density matrix becomes purely diagonal, written as
\begin{equation}
 \rho_{i} = a_{i} \lvert pp\rangle \langle pp \rvert + b_{i} 
  \lvert qq\rangle \langle qq \rvert + c_{i} \lvert pq\rangle \langle pq \rvert,
\end{equation}
\end{subequations}
where each state is given by $\lvert pp\rangle = \frac{1}{\sqrt{2}} (c_{p}^{\dagger})^{2} \lvert 0\rangle$, $\lvert qq\rangle = \frac{1}{\sqrt{2}} (c_{q}^{\dagger})^{2} \lvert 0\rangle$, and $\lvert pq\rangle = c_{p}^{\dagger} c_{q}^{\dagger} \lvert 0\rangle$, respectively.
From Eq.~\eqref{eq:rhoi} we obtain 
\begin{subequations}
\begin{equation}
\begin{pmatrix}
 a_{i} \\ b_{i} \\ c_{i}
\end{pmatrix}
 = \begin{pmatrix}
  \mathcal{R}^{2} & \mathcal{T}^{2} & 2 \mathcal{R} \mathcal{T} \\
 \mathcal{T}^{2} & \mathcal{R}^{2} & 2 \mathcal{R} \mathcal{T} \\
 2 \mathcal{R} \mathcal{T} & 2 \mathcal{R} \mathcal{T} & (\mathcal{R} - \mathcal{T})^{2}
\end{pmatrix}
\begin{pmatrix}
 a_{i-1} \\ b_{i-1} \\ c_{i-1}
\end{pmatrix},
\end{equation}
where $\mathcal{R} = \lvert r \rvert^{2} = \lvert r' \rvert^{2}$ and $\mathcal{T} = \lvert t \rvert^{2} = \lvert t' \rvert^{2}$ are the reflection and the transmission probabilities of a particle at the beam splitter, respectively.
This can be converted to the recursion relation, 
\begin{eqnarray}
 \frac{a_{i} - b_{i}}{a_{i-1} - b_{i-1}} &=& \mathcal{R}^{2} - \mathcal{T}^{2}
  , \\
 \frac{a_{i} + b_{i} - 2 c_{i}}{a_{i-1} + b_{i-1} - 2 c_{i-1}} &=& 
  (\mathcal{R} - \mathcal{T})^{2} - 2 \mathcal{R} \mathcal{T} .
\end{eqnarray}
\end{subequations}
For any beam splitter with $0 < \mathcal{R}, \mathcal{T} < 1$, the 
ratios $\lvert \mathcal{R}^{2} - \mathcal{T}^{2} \rvert$ and 
$\lvert (\mathcal{R} - \mathcal{T})^{2} - 2 \mathcal{R} \mathcal{T} \rvert$ 
are less than one. So, we obtain $a_{i} - b_{i} \to 0$ and $a_{i} + b_{i} - 2 c_{i} \to 0$ for large $i$. 
That is, the two-particle systems reach the equilibrium state ($a_{i} = b_{i} = c_{i} = 1/3$) for large $i$. Once the system reaches equilibrium, it is invariant under the scattering at the beam splitter. 
It can be shown that this property is preserved for finite temperatures, i.e., 
the two particles reach the equilibrium state at large $i$, represented by
the distribution of Eq.~\eqref{eq:P11}.

The series of beam splitters (with dephasing) is not the only way to reach 
equilibrium. Another way to achieve two-particle equilibrium and to measure the distribution, which is more suitable for massive particles (e.g. atoms), is as follows~(see Fig.~3(b)). The two particles, 
injected from two separate input ports are allowed to contact each other for 
some time until the system reaches statistical equilibrium.
To measure the distribution, the two particles should be spatially separated. The question then arises as to how the two-particle distribution is affected by the separation. This may depend on the details of the separation process. In the following, we show that any unitary process without energy exchange between the particles preserves the distribution.

The equilibrium state before separation is represented by the density matrix
\begin{equation}
 \rho_{\text{in}} = e^{-\beta H} / Z_{2},
\end{equation}
where
\begin{equation}
 H = \sum_{n} \varepsilon_{n} (c^{\dagger}_{pn} 
  c_{pn} + c^{\dagger}_{qn} c_{qn}).
\label{eq:H}
\end{equation}
The separation of the two particles can be achieved, for example, by raising the potential barrier between them~\cite{kaufman14}. This process is unitary and can be described by the scattering matrix of Eq.~\eqref{eq:ST} as long as there is no energy loss or exchange between the internal levels.
Therefore, the density matrix at the output port,  $\rho_{\text{out}}$, 
is given by
\begin{equation}
 \rho_{\text{out}} = S_{T} \rho_{\text{in}} S^{\dagger}_{T},
\end{equation}
and we find
\begin{equation}
 [H, S_T] = 0.
\end{equation}
This can be shown explicitly as follows. The Hamiltonian of Eq.~\eqref{eq:H} can be rewritten as
\begin{subequations}
\begin{equation}
H = \sum_{n} \begin{pmatrix}
c^{\dagger}_{pn} & c^{\dagger}_{qn}
\end{pmatrix}
H_{n}
\begin{pmatrix}
c_{pn} \\
c_{qn}
\end{pmatrix},
\end{equation}
where
\begin{equation}
 H_{n} = \begin{pmatrix}
\varepsilon_{n} & 0 \\
0 & \varepsilon_{n}
\end{pmatrix}.
\end{equation}
\end{subequations}
It is transformed by the scattering as 
\begin{subequations}
$H \to H' = S_{T}^{\dagger} H S_{T}$, and we find
\begin{equation}
H' = \sum_{n} \begin{pmatrix}
c^{\dagger}_{pn} & c^{\dagger}_{qn}
\end{pmatrix}
S^{\dagger} H_{n} S
\begin{pmatrix}
c_{pn} \\
c_{qn}
\end{pmatrix}.
\end{equation}
$H_{n}$ is purely diagonal, and thus we have
\begin{equation}
S^{\dagger} H_{n} S = S^{\dagger} S H_{n} = H_{n}.
\end{equation}
\end{subequations}
Therefore, $S_{T}^{\dagger} H S_{T} = H$, or $[H, S_{T}] = 0$.

Since the density matrix in the equilibrium state is a function of $H$, $\rho_{\text{in}}$ also commutes with the scattering matrix, $[\rho_{\text{in}}, S_{T}] = 0$, which leads to
\begin{equation}
 \rho_\text{out} = \rho_\text{in} S_{T} S_{T}^\dagger = \rho_\text{in}.
\label{eq:rhoout}
\end{equation}
The conclusion of Eq.~\eqref{eq:rhoout} is clear. 
The two-particle distribution is invariant with respect to the separation 
as long as the process is unitary. That is, the distributions at the exit 
ports are equivalent to the equilibrium distribution obtained in 
Eq.~\eqref{eq:P11}. In deriving this result, we have exploited
three essential properties of the system: (i) equilibrium distribution of the two particles, (ii) unitarity of the separation process, and (iii) $[S_{T}, H] = 0$. Note that our result is quite general. For an equilibrium state, the density matrix is invariant under any unitary process commuting with $H$.
This property also holds for the first setup in Fig. 3(a). Once the state reaches equilibrium, the two-particle distribution is invariant under scattering at the next beam splitter. Remarkably, this invariance of the equilibrium state is valid for any type of equilibrium involving identical particles.

%{\sf THIS part may not be necessary \\ ------
%For $T = 0$, the density matrix is reduced to
%\begin{equation}
%\rho_{\text{out}} = \rho_{\text{in}} = I / N_{g},
%\end{equation}
%where $I$ and $N_{g}$ represent the unit matrix and the ground state degeneracy, respectively. $N_{g} = 3$ and $1$ for bosons and fermions, respectively. In the case of bosons,
%\begin{equation}
%\rho_{\text{in}} = \frac{1}{3} \left( \lvert 20\rangle \langle 20 \rvert + \lvert 02\rangle \langle 02 \rvert + \lvert 11\rangle \langle 11 \rvert \right),
%\end{equation}
%and $\rho_{\text{out}}$ has the same form, 
%where each number state is given by $\lvert 20\rangle = \frac{1}{\sqrt{2}} (c_{p}^{\dagger})^{2} \lvert 0\rangle$, $\lvert 02\rangle = \frac{1}{\sqrt{2}} (c_{q}^{\dagger})^{2} \lvert 0\rangle$, and $\lvert 11\rangle = c_{p}^{\dagger} c_{q}^{\dagger} \lvert 0\rangle$, respectively.

%}

%\section{Discussion}
{\em Discussion-}.
It is rather surprising that the two-particle distribution of the type of
Eq.~\eqref{eq:P11} has not been directly verified so far. A closely related two-particle interference in the exchange scattering (known 
as the HOM effect) has been extensively studied with various particles including photons~\cite{hong87}, 
electrons~\cite{liu98,bocquillon13,freulon15}, and cold atoms~\cite{kaufman14,lopes15}, etc. 
(see, for example, Ref.~\onlinecite{bouchard20} for a review). 
%{\sf Check references...}
A direct measurement of the two-particle equilibrium distribution with the setups described in Fig.~3 would be possible with the recent remarkable progress in the realization of two-particle interference.

In the modified HOM interferometer described in Fig.~3(a), the dephasing process (between the two paths) is essential.
Without dephasing, the two particles cannot reach equilibrium, 
even if we introduce random (but unitary) scattering in the beam splitter 
array. This is because the unitary scattering itself does not change the von Neumann entropy. For instance, when $T = 0$ for two identical bosons, the initial pure state remains pure in the absence of dephasing. Introducing dephasing increases the entropy by $k \log 3$.

Two identical fermions exhibit antibunching behavior, and the equilibrium statistics show similar characteristics to the distribution measured by the standard Hong-Ou-Mandel interferometer.
In any case, due to the progress in the realization of the electronic HOM 
interferometer~\cite{liu98,bocquillon13,freulon15}, it is possible to study 
the two-fermion statistics with electrons using the setup described in 
Fig.~3(a).

Two-particle interference with cold atoms has also been 
realized~\cite{kaufman14,lopes15}. 
The type of setup shown in Figure 3(b) requires an equilibration process and is therefore suitable for massive cold atoms. 
This could be achieved, for example, by using optical tweezers to finely 
control the positions of the atoms and to tunnel between the two positions. 
Two-particle interference was demonstrated in Ref.~\onlinecite{kaufman14} 
using an oscillating distribution of $P(1,1)$, which is closely related to 
the two-particle equilibrium distribution of Eq.~\eqref{eq:P11}.

% Added on 30/4
Note that the statistics of identical particles are affected by quantum mechanics only when the two particles overlap. If the two particles never interact or overlap, quantum indistinguishability plays no role in their statistical properties (see, for example, Refs.~\onlinecite{holland95,hwang99}).

%----{\sf TO Be deleted?:
%One might ask why we need a Hong-Ou-Mandel interferometer to measure the two-particle equilibrium distribution. It would also be possible to obtain the distribution before the two particles scatter at the beam splitter. This question is valid in principle, but in order to measure the distribution, we need a spatial separation of the two particles. The point is that we cannot be sure how the separation process affects the distribution. What we have shown here is that any unitary process with no energy exchange between the two particles does not affect the distribution. The Hong-Ou-Mandel interferometer is just one example of this.
%}

%\section{}
{\em Conclusion-}.
We have studied the quantum statistics of two identical  particles. The transition between quantum and classical statistics  for two identical particles is described by considering their internal energy levels and thermal effects. Our framework shows that the effective distinguishability of the particles emerges when the thermal energy exceeds the level spacing, bridging the gap between quantum indistinguishability and classical distinguishability. The two-particle distribution function, 
$P(1,1)$, exhibits a smooth transition from purely quantum behavior at low temperatures to classical behavior at high temperatures, with a mesoscopic regime connecting the two regimes.
We have proposed experimental schemes as a realistic platform for the direct measurement of these two-particle distributions. By allowing the particles to reach statistical equilibrium prior to measurement and by exploiting the unitarity of the scattering or separation process, the equilibrium distribution is preserved, allowing a clear observation of the two-particle statistics. This approach is applicable to both bosons and fermions, with potential implementations using photons, electrons, or cold atoms.
Our results demonstrate the fundamental role of quantum mechanics in describing the emergence of classical statistics and provide a concrete method to probe this transition experimentally. 
%We believe that this study not only deepens the theoretical understanding of quantum statistics, but also opens new avenues for experimental %investigations of fundamental aspects.

%\section*{References}
%\begin{enumerate}
%    \item F. G. I: ...
%\end{enumerate}

%\begin{thebibliography}{99}
 \bibliography{references}
%\end{thebibliography}

%\newpage

%
\end{document}